# On the strong influence of inner shell resonances upon the outer shell photoionization of endohedral atoms

M. Ya. Amusia*[+1/] and L. V. Chernysheva[+]

*Racah Institute of Physics, Hebrew University, 91904 Jerusalem, Israel
[+]A. F. Ioffe Physical-Technical Institute, 194021 St.-Petersburg, Russia

**Abstract**

It is demonstrated by the example of the Xe atom stuffed inside the $C_{60}$ fullerene, i.e. the endohedral Xe@$C_{60}$, that the so-called confinement resonances in 4d subshell strongly affect the photoionization cross-section of outer 5p and subvalent 5s electrons near 4d ionization threshold.

It is a surprise that these narrow inner 4d shell resonances are not smeared out in the outer shell photoionization cross-section. On the contrary; the inner shell resonances affect the outer cross-section by enhancing them enormously.

Close to its own photoionization thresholds, 5p and 5s photoionization cross-sections of Xe@$C_{60}$ are dominated by their own confinement resonances greatly affected by the amplification of the incoming radiation intensity due to polarization by it of the $C_{60}$ electron shell. In between 4d and 5p thresholds, the effect of 4d is becoming stronger while own resonances of 5p and 5s are becoming much less important.

**1.** In this Letter, we demonstrate that a prominent oscillatory structure appears in photoionization cross-sections of 5p and 5s electrons of endohedral atom Xe@$C_{60}$ that presents Xe encapsulated inside fullerene $C_{60}$. This oscillatory structure is located mainly near and above the 4d-electrons photoionization threshold and is a consequence of very strong modification of the Xe Giant resonance under the action of the so-called confinement resonances in 4d subshell of Xe@$C_{60}$.

Soon after discovery of fullerenes $C_{60}$ in 1985, it became clear that they could be stuffed by almost all atoms A of the Periodic table and even by simple molecules. Noble gas atoms inside $C_{60}$ prefer to be at its center. Endohedrals A@$C_{60}$ are of interest from both pure scientific and application point of view – as a rather complex object with interesting electron structure and as vehicles to deliver atoms at a given place in materials and biological objects as well as building blocks of new substances (see e.g. review article [1]).

A powerful method to investigate these objects is to study their photoionization using photoelectron spectroscopy method. This method permits by fixing incoming photon and outgoing electron energies to be confident that an electron located at a given subshell of A in A@$C_{60}$ is eliminated and corresponding cross-section is investigated.

In a number of papers it was discovered, mainly theoretically, that two major effects govern the endohedral photoionization – the reflection of the outgoing electrons by the static field of the fullerenes shell [2, 3, 4] and by enhancement of the electromagnetic field at the atom A inside $C_{60}$ due to polarization of the fullerenes shell by the incoming photon beam [5]. The collision with the fullerenes shell leads to so-called *confinement resonances* [4, 2, 3] while polarization leads to *Giant endohedral resonances* [6].

The fact that encapsulating of Xe inside $C_{60}$ qualitatively modifies the Xe Giant resonance, transforming its single broad powerful maximum into four-maxima structure, was predicted in [7] and recently observed in photoionization of 4d Xe in Xe@$C_{60}^+$[8]. The theoretical results were obtained in the frame of two models one of which describes the effect of the fullerene shell as an action of a zero-thickness potential and the other takes into account also the dipole polarizability

---

of this shell by the incoming photon beam, The agreement between results in [7] and [8] encourages to make the next step and to consider how modification in the 4d Giant resonance of Xe under the action of $C_{60}$ affects the photoionization cross-section of the outer 5p and 5s electrons.

It was discovered long ago that electron correlations play a decisive role in photoionization of isolated atoms. Of particular interest is the observation that atomic Giant resonances strongly affect the photoionization cross-section of the outer subshells, leading to formation of so-called *interference* or *intershell resonances* (see [9] and references therein). It is of interest to see, how this phenomenon happens in endohedrals. In fact, the fullerenes shell can be treated as an extra multi-electron shell, thus leading to additional intershell resonances.

It appeared that the effect is very strong particularly for 5p-electrons where the direct action of fullerenes shell at the photon frequencies near and above the 4d Giant resonance is small, while for 5s it is less pronounced. It is essential also that 5p cross-section near its threshold is strongly affected by the Giant resonance of the $C_{60}$ itself, while there the role of confinement resonances proved to be inessential.

The calculations is performed in the frame of the random phase approximation with exchange (RPAE) modified in a way that permits to include the static effect of the $C_{60}$ shell via a zero-thickness potential [2] and expresses the dynamic action of $C_{60}$ upon the caged Xe via fullerenes dipole polarizability [5]. The credibility of RPAE is well established while the reasonable accuracy of the zero-thickness potential at least in obtaining qualitative predictions is confirmed by comparing the calculation [7] and measured [8] data for 4d subshell of Xe@$C_{60}$.

**2.** The photoionization amplitude of an electron from an endohedral atom's $nl$ subshell $D_{nl}^{AC}(\omega)$ in RPAE with account of reflection of photoelectrons by the $C_{60}$ shell and polarization of the latter under the action of the incoming photon beam can be presented as the following product [10]

$$D_{nl,\varepsilon l\pm 1}^{AC}(\omega) = G(\omega) F_{l\pm 1}(k) D_{nl,\varepsilon l\pm 1}^{F}(\omega). \qquad (1)$$

Here $n$ is the ionized electron principal quantum number, $l$ is its angular momentum; the polarization factor $G(\omega)$ takes into account the modification of the incoming photon beam by the fullerene $C_N$, $F_{l\pm 1}(k)$ describes the reflection factor that represent the effect of the fullerenes shell $C_N$ upon the outgoing photoelectron with the angular momentum $l\pm 1$ and linear momentum $k$, energy $\varepsilon = k^2/2$ [2/], connected to the photon frequency $\omega$ by the relation $\varepsilon = \omega - I_{nl}$, where $I_{nl}$ is the $nl$ subshell ionization potential. In (1) $D_{nl,\varepsilon l\pm 1}^{F}(\omega)$ is the atomic photoionization amplitude, in which the virtual states are modified due to action of the static potential of the fullerenes shell upon the virtually excited atomic states.

Taking into account that the atom's A radius $R_A$ is considerably smaller than the fullerenes radius $R_C$, a rather simple expression can be obtained for $G(\omega)$

$$G(\omega) = 1 - \frac{\alpha_C(\omega)}{R_C^3}. \qquad (2)$$

Here $\alpha_C(\omega)$ is the dipole polarizability of the fullerenes shell. Formula (2) was derived in [5] under simplifying assumption that $R_A/R_C \ll 1$. While $\alpha_C(\omega)$ is difficult to calculate ab-

---

[2/] Atomic system of units with electron charge $e$, mass $m$, and Plank constant $\hbar$ being equal to 1, $e = m = \hbar = 1$.

initio, it can be easily expressed via experimentally quite well known photoionization cross-section $\sigma_C(\omega)$ of the $C_{60}$ [11] and references therein):

$$\operatorname{Re}\alpha_C(\omega) = \frac{c}{2\pi^2}\int_{I_F}^{\infty}\frac{\sigma_C(\omega')d\omega'}{\omega'^2-\omega^2}.$$

$$\operatorname{Im}\alpha_C(\omega) = c\sigma_C(\omega)/4\pi\omega$$

(3)

Here $I_C$ is the fullerene ionization potential and $c$ is the speed of light.

Since the cross-section $\sigma_C(\omega)$ is absolutely dominated by the fullerenes Giant resonance that have a maximum at about 2Ry, $G(\omega)$ starts to decreases rapidly at $\omega > 2Ry$ reaching its asymptotic value equal to 1 at about 5Ry. This factor, connecting the atomic and fullerenes photoionization cross-section, is able to alter the endohedral cross-section as compared to pure atomic one.

To obtain the reflection factor $F_{l\pm1}(k)$, we substitute the fullerenes shell action by a static zero-thickness potential [2]

$$W(r) = -W_0\delta(r-R).$$

(4)

The parameter $W_0$ is defined from the condition that the binding energy of extra electron in negative ion $C_{60}^-$ is equal to the experimentally observed value. The factor $F_{l'}(k)$ is determined by the expression [2, 3, 10]:

$$F_l(k) = \cos\Delta\delta_l(k)\left[1-\tan\Delta\delta_l(k)\frac{v_{kl}(R)}{u_{kl}(R)}\right],$$

(5)

where $\Delta\delta_l(k)$ is the addition to the photoelectron elastic scattering phase of the partial wave $l$ due to action of the potential (4), $u_{kl}(r)$ is the regular and $v_{kl}(r)$ irregular at point $r \to 0$ radial parts of atomic Hartree-Fock one-electron wave functions. The following relation expresses the additional phase shift $\Delta\delta_l(k)$:

$$\tan\Delta\delta_l(k) = \frac{u_{kl}^2(R_C)}{u_{kl}(R_C)v_{kl}(R_C)-k/2W_0}.$$

(6)

The factor $F_{l'}(k)$ as a function of $k$ oscillates due to interference between the direct photoelectron wave and its reflections from the fullerenes shell. This factor redistributes the resulting cross section as compared to that of the isolated atom but cannot change its value integrated over essential $\omega$ region.

We obtain $D^F_{nl,\varepsilon l\pm1}(\omega)$ in the frame of the RPAE. When the fullerene shell is presented by the potential (4), the following equation [10, 12]:

$$\langle \varepsilon l' | D^F(\omega) | nl \rangle = \langle \varepsilon l' | d | nl \rangle +$$

$$+ \sum_{\varepsilon''l'',\varepsilon'''l'''} \frac{\langle \varepsilon'''l''' | D^F(\omega) | \varepsilon''l'' \rangle \left[ F^2_{\varepsilon'''l'''} n_{\varepsilon''l''}(1-n_{\varepsilon'''l'''}) - F^2_{\varepsilon''l''} n_{\varepsilon'''l'''}(1-n_{\varepsilon''l''}) \right]}{\omega - \varepsilon_{\varepsilon'''l'''} + \varepsilon_{\varepsilon''l''} + i\eta(1 - 2n_{\varepsilon''l''})} \times \quad (7)$$

$$\times \langle \varepsilon''l'', \varepsilon l' | U | \varepsilon'''l''', nl \rangle.$$

Here $\langle \varepsilon l' | D^F(\omega) | nl \rangle \equiv D^F_{nl,\varepsilon l \pm 1}(\omega)$, $F_{\varepsilon'l'} \equiv F_{l'}(k')$, $d$ is the one-electron operator that describes photon-electron interaction in the dipole approximation; summation over vacant levels includes integration over continuous spectrum, $n_{\varepsilon l}$ is the Fermi step function that is equal to 1 for $nl \leq F$ and 0 for $nl > F$, where $\leq F (> F)$ denotes occupied (vacant) atomic levels in the target atom; $\eta \to +0$ and the Coulomb interelectron interaction matrix element is defined as $\langle \varepsilon''l'', \varepsilon l' | U | \varepsilon'''l''', nl \rangle = \langle \varepsilon''l'', \varepsilon l' | r_</r_>^2 | \varepsilon'''l''', nl \rangle - \langle \varepsilon''l'', \varepsilon l' | r_</r_>^2 | nl, \varepsilon'''l''' \rangle$. In the latter formula notation of smaller (bigger) radiuses of $r_<(r_>)$ of interacting electron coordinates comes from the well-known expansion of the Coulomb interelectron interaction. The necessary details about solving (7) one can find in [9, 12].

The cross section $\sigma^{AC}_{nl}(\omega)$ for $nl$-subshell photoionization is expressed via the partial amplitudes (1) using the following relation:

$$\sigma^{AC}_{nl}(\omega) = 5.738\omega [l | D^{AC}_{nl, \varepsilon l-1} |^2 + (l+1) | D^{AC}_{nl, \varepsilon l+1} |^2] Mb, \quad (8)$$

**3.** In our calculations $nl = 5p; 5s$. As intermediate states in (7) as occupied we included the following states $n''l'' = 5p; 5s, 4d$. However, the main contribution in the considered frequency range comes from $4d - \varepsilon f, \varepsilon p$ transitions.

The equation (7) differs from RPAE equation for isolated atoms by the presence of factors $F_{\varepsilon'l'}$ in the sums and in the integrand. So, at first glance their oscillations should be integrated over. It does not happen, however.

The results of calculations are presented in Fig. 1-3. Isolated atom cross-section is marked as "Xe free, RPAE". If in the amplitude (1) $G(\omega)$ is neglected and $D^F_{nl,\varepsilon l\pm 1}(\omega)$ is substituted by $D_{nl,\varepsilon l\pm 1}(\omega)$ - solutions of (7) without factors $F_{\varepsilon'l'}$ under the sum, the cross-section marked "Xe@C$_{60}$, RPAE" is obtained. Neglecting $G(\omega)$ we obtain from (7) the cross-section that is marked "Xe@C$_{60}$, FRPAE". Using complete amplitude given by (1), we obtain the cross-section that is marked "Xe@C$_{60}$, GFRPAE".

Fig. 1 displays the photoionization cross-section of 5p-electrons in Xe and Xe@C$_{60}$ near and above the 4d threshold, from $\omega = 4.5 Ry$ till $\omega = 9 Ry$. Prominent oscillations of endohedral cross-section relative to its values in isolated atom appear. Comparison of "Xe@C$_{60}$, GFRPAE" with "Xe@C$_{60}$, RPAE" cross-sections leaves no doubt that the effect is due to inclusion of reflection of photoelectron's from the intermediate 4d-subshell excitations. It is seen also that the effect of the polarization factor $G(\omega)$ completely disappears at $\omega > 5 Ry$.

Fig. 2 depicts the cross-section of 5p-electrons in Xe@C$_{60}$ above the 5p threshold and up to $\omega = 4 Ry$ with account of photoelectrons reflection and photon beam modification. Data for isolated Xe are also presented. Here, contrary to Fig.1, close to 5p ionization threshold a profound maximum appeared called Giant endohedral resonance [6]. The important influence of $G(\omega)$ is seen up to $\omega = 4 Ry$.

Fig. 3 presents the photoionization cross-section of 5s-electrons in Xe@C$_{60}$ near and above the 4d threshold. We see strong modifications due to 4d confinement resonances. The cross-section is, however, much smaller than in Fig. 1 and the difference between "Xe@$C_{60}$, GFRPAE" and "Xe@$C_{60}$, FRPAE" on one side and "Xe@$C_{60}$, RPAE" cross-sections on the other is less pronounced than for 5p-electron, given in Fig. 1. It is interesting that the effect of polarization factor $G(\omega)$ leads to a powerful resonance very close to 5s threshold – at about 2Ry. It enhances essential also a maximum at 4Ry.

Results for photoionization of 5s-electrons of Xe@$C_{60}$ obtained in [13] are presented also. There the finite-width instead of zero thickness potential (4) was used to represent the fullerenes shell action. It leads to considerably weaker and less pronounced structure. Although the finite width potential seems to be more realistic than (4) for 5p and 5s electrons in Xe@$C_{60}$, special efforts must be undertaken [14] to reproduce with its help [8]. This is why we do believe that using (4) is not only simple but more reliable.

**4.** We performed calculations of the photoionization cross-section of 5p and 5s electrons in the endohedral Xe@$C_{60}$. We found that confinement resonances in virtual excitations of 4d Giant resonance affect strongly the cross-sections for transitions from *5p* and *5s* subshells. The results demonstrate extremely powerful effect played by confinement resonances in 4d upon 5p and 5s photoionization. At the same time, in shaping the resonance structure of the 5p and 5s cross-sections the polarization factor is of importance.

The information that could come from studies of 5p and 5s cross-sections is of great interest and value. Thus, the experimental search for the predicted here cross-sections although being very complicated is desirable and promising.

**Acknowledgement**

The authors acknowledge the support received from the Israeli-Russian Grant RFBR-MSTI no. 11-02-92484.

**References**
1. V. K. Dolmatov, in Theory of Confined Quantum Systems: Part Two, edited by J. R. Sabin and E. Brändas, Advances in Quantum Chemistry (Academic Press, New York, 2009), Vol. **58**, 13 (2009).
2. M. Ya. Amusia, A. S. Baltenkov, and B. G. Krakov, Phys. Lett. A, **243**, p. 99-105 (1998).
3. A. S. Baltenkov, J. Phys. B: At. Mol. Opt. Phys. **38**, L169 (1999).
4. J.-P. Connerade, V. K. Dolmatov, and S. T. Manson, J. Phys B: At. Mol. Opt. Phys. **33**, 2279 (2000).
5. M. Ya. Amusia and A. S. Baltenkov, Phys. Rev. A **73**, 062723 (2006).
6. M. Ya. Amusia, A. S. Baltenkov and L. V. Chernysheva, JETP Letters, **87**, 4, 230-233, 2008.
7. M. Ya. Amusia, A. S. Baltenkov, L. V. Chernysheva, Z. Felfi, and A. Z. Msezane, J. Phys.: At. Mol. Opt. Phys. B **38**, L169 (2005).
8. A. L. D. Kilcoyne, A. Aguilar, A. M¨oller, S. Schippers, C. Cisneros, G. Alna'Washi, N. B. Aryal, K.K. Baral, D. A. Esteves, C. M. Thomas, and R. A. Phaneuf, Phys. Rev. Lett. **105**, 213001 (2010).
9. M. Ya. Amusia, Atomic Photoeffect (New York: Plenum Press) (1990).
10. M. Ya. Amusia, Simple and Onion-type Fullerenes shells as resonators and amplifiers, Fullerene, Nanotubes, and Carbon Nanostructures, **18**, 4-6, 353 – 368 (2010).
11. J. Berkowitz, J. Chem. Phys. **111**, 1446 (1999).
12. M. Ya. Amusia, L. V. Chernysheva and V. G. Yarzhemsky, Handbook of theoretical Atomic Physics, Data for photon absorption, electron scattering, and vacancies decay, Springer, Berlin, in print (2012).
13. V. K. Dolmatov and S. T. Manson, J. Phys.: At. Mol. Opt. Phys. B **41**, 165001 (2008).
14. V. K. Dolmatov and D. A. Keating, J. Phys.: Conf. Series (accepted); e-print arXiv:1109.5292.

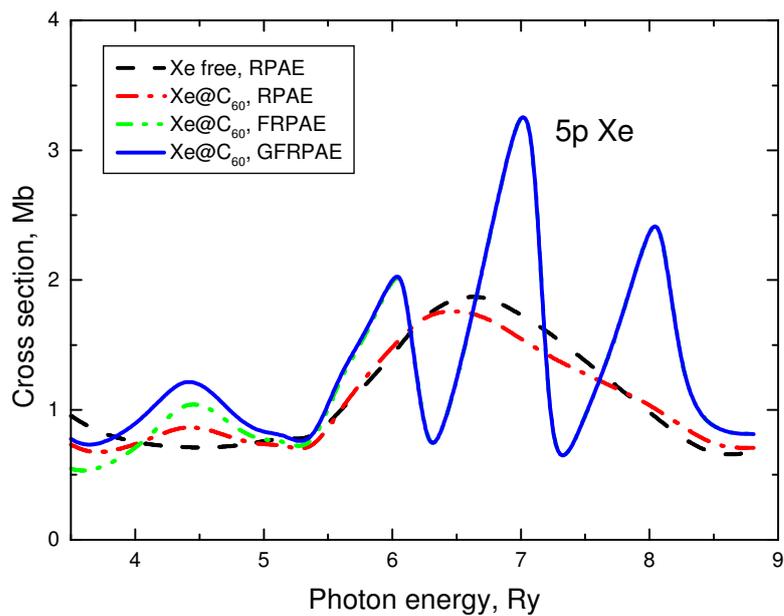

Fig. 1. Photoionization cross-section of 5p-electrons in Xe@$C_{60}$ near and above the 4d threshold with account of photoelectrons reflection and photon beam modification Data for isolated Xe are also presented.

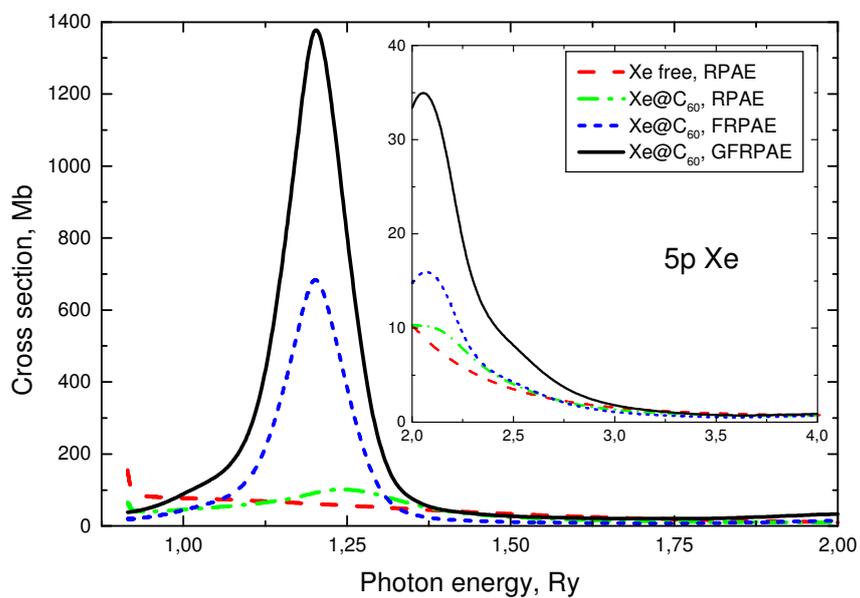

Fig. 2. Photoionization cross-section of 5p-electrons in Xe@$C_{60}$ above the 5p threshold with account of photoelectrons reflection and photon beam modification. Data for isolated Xe are also presented.

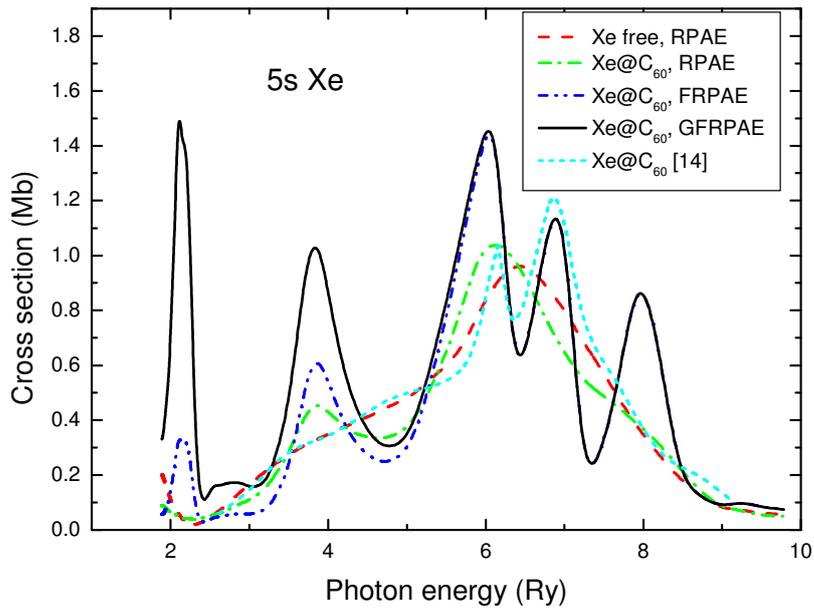

Fig. 3. Photoionization cross-section of 5s-electrons in Xe@$C_{60}$ near and above the 4d threshold with account of photoelectrons reflection and photon beam modification. Results from previous calculation [14] are given. Data for isolated Xe are also presented.